\documentclass[twocolumn,showpacs,preprintnumbers,amsmath,amssymb]{revtex4}
\usepackage{latexsym}
\usepackage{graphicx}
\usepackage{amsmath}

\begin{document}

\bibliographystyle{apsrev}
\newtheorem{theorem}{Theorem}
\newtheorem{corollary}{Corollary}
\newtheorem{definition}{Definition}
\newtheorem{proposition}{Proposition}
\newtheorem{lemma}{Lemma}
\newcommand{\ignore}[1]{}
\newcommand{\proofend}{\hfill\fbox\\\medskip }
\newcommand{\proof}[1]{{\bf Proof.} #1 $\proofend$}
\newcommand{\nn}{{\mathbbm{N}}}
\newcommand{\rr}{{\mathbbm{R}}}
\newcommand{\cc}{{\mathbbm{C}}}
\newcommand{\mbp}{\ensuremath{\spadesuit}}
\newcommand{\je}{\ensuremath{\heartsuit}}
\newcommand{\jd}{\ensuremath{\clubsuit}}
\newcommand{\id}{{\mathbbm{1}}}
\renewcommand{\vec}[1]{\boldsymbol{#1}}
\newcommand{\me}{\mathrm{e}}
\newcommand{\mi}{\mathrm{i}}
\newcommand{\md}{\mathrm{d}}
\newcommand{\sg}{\text{sgn}}

\delimitershortfall=-2pt

\title{Quantum coherence in ion channels: Resonances, Transport and Verification}
\author{Alipasha Vaziri}
\affiliation{Janelia Farm Research Campus, Howard Hughes Medical Institute 19700 Helix Drive Ashburn, VA 20147, USA}

\author{Martin~B.~Plenio}
\affiliation{Institut f\"ur Theoretische Physik, Albert-Einstein
Allee 11, Universit\"at Ulm, Germany}
%

\begin{abstract}
Recently it was demonstrated that long-lived quantum coherence
exists during excitation energy transport in photosynthesis.
It is a valid question up to which length, time and mass scales
quantum coherence may extend, how to one may detect this coherence
and what if any role it plays for the dynamics of the system.
Here we suggest that the selectivity filter of ion channels may exhibit quantum coherence which might be relevant for the process of ion selectivity and conduction. We show that quantum resonances could provide an alternative approch to ultrafast 2D spectroscopy to probe these quantum coherences. We demonstrate that the emergence of resonances in the conduction of ion channels that are modulated periodicallly by time dependent external electric fields can serve as signitures of quantum coherence in such a system. Assessments of experimental feasibility and specific paths towards the experimental realization of such experiments are presented.
We show that this may be probed by direct 2-D spectroscopy
or through the emergence of resonances in the conduction of ion
channels that are modulated periodically by time dependent
external electric fields.
\end{abstract}

\maketitle

\date{\today}

A wide variety of transport processes on all length and time-scales
are of fundamental importance for the function of biological systems.
For a long time it was assumed that these transport processes may be
modelled accurately as classical random processes and the possible
relevance of non-classical phenomena such as quantum coherence was
mostly ignored.
The experimental demonstration of the presence of long-lived quantum
coherence in the photosynthetic energy transfer at cryogenic
temperatures \cite{Engel07,Marangos} and room temperature \cite{Engel10}
has forced a critical reevaluation of this position. As a consequence,
the detailed experimental and theoretical study of quantum coherence
in bio-molecular systems and its possible relevance to explain functional
properties of those systems is currently receiving rapidly increasing
attention.

The transport in the photosynthetic complex or the reactions during
the first steps in vision \cite{schoenlein91,wang94}
involve initiation and fast transfer of
electronic excitation via an absorbed photon on the femtosecond time
scale. Given that these systems have been evolutionary optimized
to function at highest efficiencies (e.g. human vision has few
photon sensitivity \cite{Hechtvision,singlephotonvision}),
they might seem to naturally suggest themselves as areas to look
for quantum coherence. More generally however, on time scales that
are short compared to the decoherence time of an
observed biological process quantum mechanical coherence
can be expected to occur and play a role in biological
function despite (or even because of) the strong interaction
of bio-molecular systems with their environment.

Indeed, there may be non-trivial effects, positive or negative, due
to the interplay between quantum coherent dynamics and decoherence
from environments. In models of photosynthetic energy transfer it
has been shown that compared to the purely quantum coherent case the
efficiency of excitation transfer is enhanced by the presence of
dephasing noise \cite{Aspuru08,PlenioH08} originating from an
environment of vibrational modes. It was shown that the transport
performance of the system is optimal when it is neither fully
decohered and hence classical nor fully quantum coherent. The
mechanisms underlying this phenomenon have been identified and
understood to apply more generally beyond the specific
setting of energy transport in photosynthetic complexes
\cite{CarusoCDHP09,ChinDCHP09} and optimized transport settings
in simpler structures have been determined \cite{CaoS09}.
The role of quantum coherence in other light harvesting systems
\cite{Olaya} and for speed-up at early stages of the transport
process has been examined \cite{Whaley} and the presence of
entanglement \cite{CarusoCDHP09,BioEntanglement} has been studied in some detail.

These results provide further motivation to explore the relevance of
quantum coherence and decoherence for transport in other biological
entities. Here we discuss another system in which the interplay between quantum
coherence and environmental noise could be responsible to explain
some of the functional observations of the system.

Ion-channels are protein complexes that regulate the flow of
particular ions across the cell membrane and are essential for
a large range of cellular functions \cite{Hille2001}. Besides
their role in neuronal communications in which voltage gated
channels and ligand gated channels are involved in the generation
of action potentials and mediating synaptic release, more generally
ion-channels play a key role in processes that rely on fast responses
on the bio-molecular scale. Examples include muscle contraction,
epithelial transport and T-cell activation \cite{Hille2001, Triggle2006}.
Structurally, ion-channels are
integral transmembrane protein complexes with multiple subunits
whose relative spatial arrangement forms a pore through which
ions can flow in or out of the cell along a concentration
gradient. A common feature in a large number of ion-channels
is the presence of a gate that can be activated by chemicals,
voltage, light or mechanical stress and a selectivity filter
which is responsible for allowing only a specific type of ions
to pass. The selectivity filter is only a few angstroms wide
hence the ions have to move through it in a single file fashion
without their hydration shell \cite{RouxSchulten}.

A large number of ion-channels have been successfully crystallized
over the last decade and x-ray crystallographic data have provided
us with structural understanding of these protein complexes on the
atomic level \cite{Doyle98, Dutzler2002, Sokolova2001, Wang2004}.
In parallel based on this data and by using molecular dynamics
simulation different functional models for the
mechanisms of ion selectivity and conduction have been developed
\cite{Noskov2004, Berneche2001, Dmitriev2006, Garafoli2003, Gwan2007},
however experimental observation of these dynamics in real
time and on atomic level has remained challenging.

Amongst the best studied ion-channels are potassium channels. Their
family includes a wide range of channels with different conductivities
and gating mechanism including the voltage gated potassium channels
which are responsible for restoring the membrane potential during the
course of action potentials in neurons \cite{Hille2001}. While the gating
mechanism can greatly vary across different types of potassium
channels, the sequence of amino acids forming the selectivity
filter is fairly conserved across all potassium channels. The
structure of the selectivity filter has been best studied in
the bacterial KcsA channel \cite{Doyle98} on which the following
discussions are focused.

During its open gate state an efflux of $10^{8}$ ions per second can
be experimentally observed \cite{Gouaux2005} which is close to the
diffusion limited rate. At the same time the selectivity filter is
capable of selecting potassium over sodium with a ratio of at least
$10^4:1$ \cite{Doyle98}. Furthermore, it is known from
the x-ray crystallographic data of the KcsA that its selectivity filter
forms a pore with the diameter of $\sim 0.3$ nm and a length of $\sim 1.2$ nm.
This implies that individual potassium ions cannot move through
the selectivity filter with their 8 water molecule hydrated
shell which must be shed before entering the selectivity filter.
The transfer through the filter then proceeds in single file
fashion where potassium and water molecules alternate. This
makes the mentioned observations on transmission and discrimination
rates even more remarkable.
In a number of models of the selectivity filter this fact is
accounted for by assuming that the ion transport is achieved via
an interplay of ion - filter attraction and ion - ion repulsion
while the selectivity is achieved through a mechanism which
exploits the lower water shell dehydration energy for potassium
than for sodium \cite{Morais-Cabral2001, Gouaux2005, Noskov2004}.
Certainly, the magnitude of the thermal fluctuations
of the backbone atoms forming the selectivity filter is large relative
to the small size difference between Na and K, raising fundamental
questions about the mechanism that gives rise to ion selectivity.
This suggests that the traditional explanation of ionic selectivity
should be reexamined \cite{RouxSchulten}.

\begin{figure}[t]
\includegraphics[width=8.5cm]{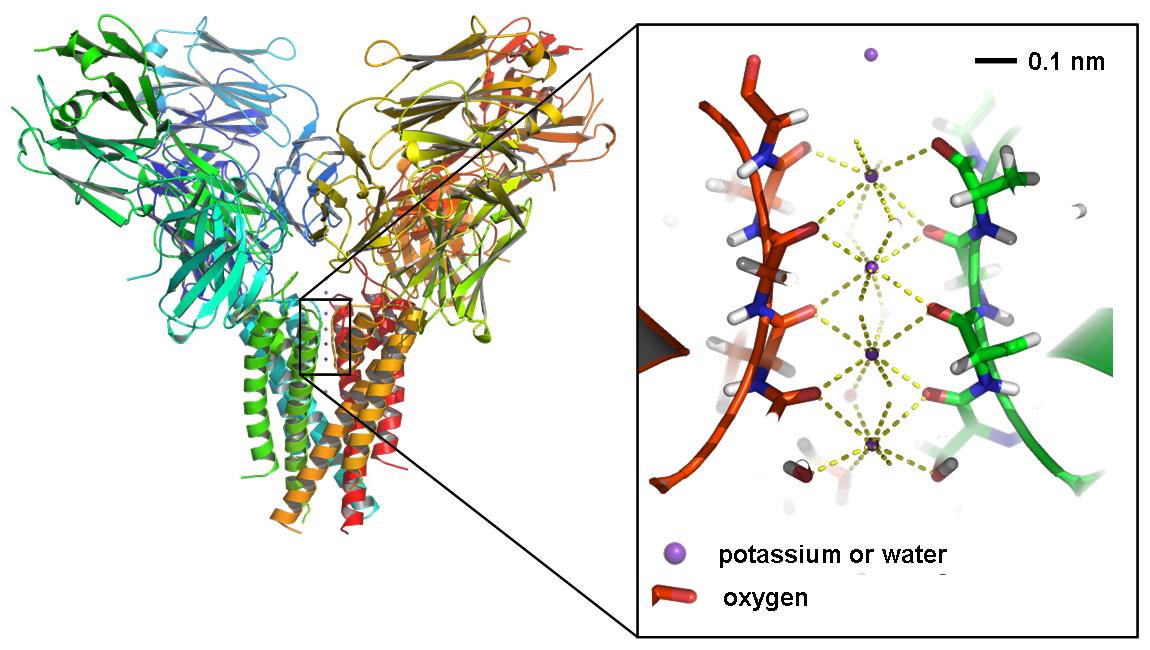}
\caption{\label{KcsA}Schematic illustration of the KcsA postassium
channel after PDB 1K4C. KcsA protein complex with four
transmembrane subunits (left) and the selectivity with four
axial trapping sites formed by the carbonyl oxygen atoms in
which a potassium ion or a water molecule can be trapped.}
\end{figure}

A closer look at the involved dimensions and energetics of the
process reveals that the underlying mechanism for ion transmission and
selectivity might be not entirely classical. Fig.\ref{KcsA} shows the
structure of the KcsA and the details of the selectivity filter. It is
formed by four chains of amino acids each from one of the
four protein complex subunits constituting the ion channel.
Each chain is made of five amino acids from whom a carbonyl
group oxygen atom is pointing toward the pore (Fig.\ref{KcsA} inset).
This configuration leads to a series of axial potential minima
(i.e. binding sites) and maxima for the potassium ions. In each
of these binding sites which are formed by four in plane oxygen
atoms either a dehydrated potassium ion or a water molecule can
be trapped (Fig.\ref{KcsA} inset). Numerical calculations of free
energy maps of the ion conduction shows two energetically nearly
degenerate configuration paths for the ion transport which are
separated by only $\sim 2$kcal/mol \cite{Berneche2001}. The transport
along each
configuration path involves a coordinated dislocation of
water -potassium chain. During these transitions different
sequences of water and potassium occupy the four binding sites.
This is also supported by the x-ray diffraction data which has
revealed two equally represented populations corresponding to
two different sequences of water and potassium in the filter
which are referred to as 1,3 and 2,4 state \cite{Morais-Cabral2001}.
In this
arrangement the axial separation of the potential minima is
$\sim 0.24$ nm and the height of the barrier fluctuates based
on the presence or absence of an ion at a particular site and
the thermal vibrations of the protein in the range of
$\sim 1-5$ kcal/mol corresponding to $\sim 1.7-8 k_{B}T$
\cite{Gwan2007}. This
situation allows for at least two different but equivalent
considerations through which quantum coherence could be in
involved in the ion transmission process.

First, based on the thermal energy of $E = k_{B}T/2 \approx
2\times 10^{-21} J$ of the ions the de-Broglie wavelength
$\lambda_{DB} = h/\sqrt{2mE}$  of the potassium matter wave
can be estimated to be $\sim$0.05 nm. Given that the periodicity
of the axial modulation of the potential is $\sim$0.25nm it is
within a factor of five of the potassium matter wavelength. In
this sense the transmission could involve a process which can
be viewed as the diffraction of the potassium matter wave off
a one dimensional axial "grating" formed by the modulations of
the potential energy.

Alternatively the transmission could be seen to involve quantum
tunneling through the potential barrier between individual
neighboring binding sites. This can seen by estimating the
tunneling probability, $p_{tun} = e^{-\Delta\sqrt{2m(E_{0}-E)/h^{2}}}$,
with $E$ the energy of a potassium ion, $m$, the mass of potassium,
$E_0-E$ the height and $\Delta$ the width of the tunneling
barrier. Using the parameters of the selectivity filter
\cite{Gwan2007, Berneche2001} this yields a tunneling probability
of $\sim 7\times10^{-5}$ for a single "try". Multiplying this value
with the trapping frequency of a bound ion in a harmonic potential
of $\sim 10^{12}Hz$ yields an overall transmission approaching the
observed rates of ~$10^8$ ions per second. This
estimate assumes a square well potential. However, a more accurate
description would assume a quartic double well potential
$V(x) = \alpha x^4-2\beta x^2$ whose parameters are chosen such
that the potential barrier at $V(0)$ has a height $\Delta E = 0.04eV = 1.5kT$,
and a separation of the minima of $\sim 0.24nm$. Furthermore,
assuming that the degree of excitation inside a potential well
will be compatible with the thermal energy, we can then compute
the energy level splitting between levels that are split due to tunneling.
This splitting gives the tunneling rate for a system
in that energy level. Typical values range between $\sim\! 10^5 s^{-1}$
and $\sim\! 10^{10}s^{-1}$ for energy levels around the thermal
energy. In these estimations we have assumed a static potential
and have ignored any rotational vibration modes in the carbonyl
groups. However, it has been theoretically shown \cite{Gwan2007,IonChannels}
that these could significantly lower the
potential barriers, so that the tunneling probabilities given
above would this sense represent only a lower bound.
%

Based on these estimations we hypothesize that quantum coherence
could be present in the selectivity filter and might be involved
in explaining some of the functional features such as the high
ion conduction rate and ion discrimination rate. However, given
the strong interaction of the system with the environment these
coherences may be short lived. A rough estimate suggests that
the decoherence time will be on the order of $10^{-9} - 10^{-8}s$
and thus of the order of the expected tunneling rates \cite{Estimate}.
In fact, we expect that the interplay between quantum coherent dynamics
and decoherence from the environment might actually be necessary
for explaining the dynamics of the selectivity filter.
As demonstrated for the excitation transport through photosynthetic
complexes the presence of dephasing noise \cite{Aspuru08,PlenioH08}
may even enhance the efficiency of transport compared to the purely
quantum coherent case.

A key feature of such coupled quantum systems that we will be
exploiting here is that interference, constructive or destructive,
in the underlying dynamics may lead to resonances in the transport
efficiency. These resonances allow us to infer the presence of
quantum coherence as they are absent in pure rate equation models.
These resonances may be measured directly in the energy, particle
or charge current that the systems allows -- the measurement of
correlations in time or coherence in relation to the driving field
are not required.

This represents two distinct advantage. First the direct detection
and verification of quantum coherence in biological systems has
been so far been based mainly on two (or higher) dimensional
femtosecond spectroscopy techniques \cite{Hochstrasser2007} in which
quantum coherence manifests itself as a beating signal in the intensity
of the cross peaks. In two-dimensional femtosecond spectroscopy three
pulses and a strongly attenuated local oscillator are incident on the
sample. The first pulse creates a coherence that evolves for a time period
usually referred to as the evolution or coherence time, then the second
pulse creates an excited-state population that evolves over a second
time period called the waiting time or mixing period. Finally, the third
pulse generates a coherence that accumulates phase in the opposite
direction. Thereafter re-phasing occurs and a signal pulse is emitted in
a direction determined by the phase matching condition. The coherence
time and the detection time are scanned in a range of a few hundred
picoseconds in the vibrational 2D spectroscopy and in the range of
femtoseconds for studies in the visible range while the signal is
measured through a heterodyne detection for each combination of the
coherence and the detection time. In this way coupling strengths
and energy transfer in molecules are probed. In the absence of coupling contributions from the excited-state absorption and emission cancel each
other and as a result no off-diagonal peaks appear in the spectrum. In the presence of coupling the contributions from the excited state absorption and emission do not fully cancel and a so-called "cross peak"
emerges in the spectrum. Intuitively speaking, the 2D spectroscopy can probe the
"memory" of a system. As a result it can be used to examine how fast
an initially coherent state decoheres with time.

Even though several laboratories are now mastering this method in the visible
\cite{Engel07,Engel10,Scholes,Marangos}and in the IR regime \cite{Hamm98, Khalil2000, Demirdoven2002, Finkelstein2007, Tokmakoff2007, Brendenbeck2007} its implementation remains an experimental challenge.
Besides the usual complexity involved in the design, operation and
maintenance of a coherent ultrafast 2D spectroscopy
system, it implies stringent requirements for stability and
precision for the inter-pulse delay times.


Second, in some systems, such as in current-carrying polymers
or in the selectivity filter the observation of resonances
in transport rates  may not only provide an alternative approach
towards the verification of the existence of quantum coherence
but also a means for direct demonstration of its importance
for the system dynamics. Indeed, a further potentially significant
implication of our work is that the presence of quantum coherence
for example in the selectivity filter and the concomitant existence
of sharp resonances in the transport may point towards a novel
mechanism for the explanation of the high degree of selectivity.
Indeed, relatively small changes in the system parameters (mass,
ion radius, interaction strengths of ions traversing the channel)
may then lead to sharp changes in the conductance.

{\em Quantum coherence \& transport --}
The dynamics of ion conduction inside an ion channel
is highly complex due to the manifold interactions between the ions
and the multiple degrees of freedom of the channel. In the following
quantum mechanical analysis of a transport process it is our aim
to avoid this complexity while retaining some structural elements
of the dynamics of an ion channel. This will allow us to show that
in this simplified model quantum coherence can give rise to very
significant effects in the conductance of the channel. In turn
this suggests that a more realistic model may also exhibit similar
features. This provides the motivation for considering possible
experimental set-ups that would be able to verify or falsify the
hypothesis of the presence of coherence in a real ion channel.

The theoretical and numerical discussion, first defines the
basic dynamical equations governing the transport and discusses
their relationship to real systems. Then we move on to demonstrate
that coherently driven transport channels exhibit resonances in
their conductance which, in turn, can be linked to the level
of quantum coherence in the system. In fact, we will elucidate
the mechanism behind these resonances and hence understand why
they do not occur in systems exhibiting classical rate equation
dynamics. Based on the insights that we gain from this analysis
we then proceed to discuss in some detail the experimental
feasibility of our predictions and discuss the concrete settings
underlying our current experimental efforts in this direction.

{\em The basic dynamical equations --} In the present context we
are focusing on quantum dynamical equations that are motivated
by the example of the selectivity filter of ion channels. Nevertheless,
their essential features should capture a wide variety of transport
processes in bio-molecular systems. One of the underlying assumption
is that to a good approximation the transport is particle number
conserving, that is, it is described by a tight-binding type
Hamiltonian that is mainly subject to
dephasing-type noise. This is a reasonable model assumption for
excitation transport in photosynthesis (excitation annihilation
rates are short compared to the observed transport times) or the
transport of ions through an ion channel (as ions rarely escape
sideways through the channel walls). As we are principally motivated
by ion channels we restrict attention to a linear chain of sites
in which an excitation may be exchanged between nearest neighbors
even though the ideas presented here are not limited to that setting.
Note that an excitation at a particular site in
the following Hamiltonian model does not need to represent the
presence of a potassium ion at a certain position in space. Rather
each site may also be thought of as specific potassium-water
single file configurations that may arise during the transport. Each
of these
configurations will sit in an effective potential well and is separated
from other configurations by a potential barrier that it may traverse
through quantum coherent tunneling or thermal activation
\cite{RouxSchulten}. The coherent
part of the evolution
is then described by the Hamiltonian
\begin{equation}
        H/\hbar = \sum_{k=1}^N \omega_k(t) \sigma_k^{+}\sigma_k^{-} +
        \sum_{k=1}^{N-1} c_k(t) (\sigma_k^{+}\sigma_{k+1}^{-}
        + \sigma_k^{-}\sigma_{k+1}^{+})
\end{equation}
with hopping rates $c_k(t)$ and site energies $\hbar\omega_k(t)$ that
may vary across sites and in time. Let us furthermore assume that
the first and last site of the chain are connected to environments
that insert or remove excitations from the system or, in other words,
drive transitions between configurations. The first site
may for example be linked to a source that supplies particles
(electrons from a lead or potassium ions from the cytoplasm entering
the ion channel thus driving the system to a different configuration)
while the last site may then be linked to a similar
environment but with lower voltage and/or lower concentration of
particles. In general one may describe such processes by Lindblad
\cite{LindbladIntroduction}
operators of the type
\begin{eqnarray}
        {\cal L}_{s/d}(\rho) &=& \Gamma_{s/d}(n_{s/d}+1) [ -
        \sigma_1^{+}\sigma_{1}^{-}\rho -\rho\sigma_1^{+}\sigma_{1}^{-}
        + 2 \sigma_{1}^{-}\rho\sigma_{1}^{+}] \nonumber\\
        &&+ \Gamma_{s/d}n_{s/d} [ -\sigma_1^{-}\sigma_{1}^{+}\rho
        - \rho \sigma_1^{-}\sigma_{1}^{+}
        + 2 \sigma_{1}^{+}\rho\sigma_{1}^{-}]
\end{eqnarray}
where s (d) stand for source (drain), $\Gamma_{s}$ ($\Gamma_{d}$)
for the coupling strength between the chain and the environments
and where $n_s$ ($n_d$) represent concentrations. Terms of the form
$\sigma_1^{+}\sigma_{1}^{-}\rho -\rho\sigma_1^{+}\sigma_{1}^{-}$
are responsible for the evolution of the system when no excitation
has entered or left the system, while the term
$2 \sigma_{1}^{-}\rho\sigma_{1}^{+}$ ($2 \sigma_{1}^{-}\rho\sigma_{1}^{+}$)
describes the event when an excitation has left (entered) the system.

Dephasing noise also originates from an interaction with an
environment, for example with the oscillatory degree of freedom
of the carbonyl groups in the ions channel, but does not create or
destroy excitations in the chain. Dephasing can be described
by another Lindblad operator of the form
\begin{equation}
        {\cal L}_{deph}(\rho) = \sum_k \gamma_{k} [ -\{
        \sigma_k^{+}\sigma_{k}^{-},\rho
        \}
        + 2 \sigma_{k}^{+}\sigma_{k}^{-}\rho\sigma_{k}^{+}
        \sigma_{k}^{-}]
\end{equation}
with site dependent dephasing rates $\gamma_k$. It should be
noted that this model ignores temporal and spatial correlations
in the noise. While such correlated noise may lead to quantitative
changes in the following considerations the basic principle
of using the presence or absence of resonances in the conductivity
to determine the presence of quantum coherence remains valid.
\begin{figure}[h]
\includegraphics[width=8.3cm]{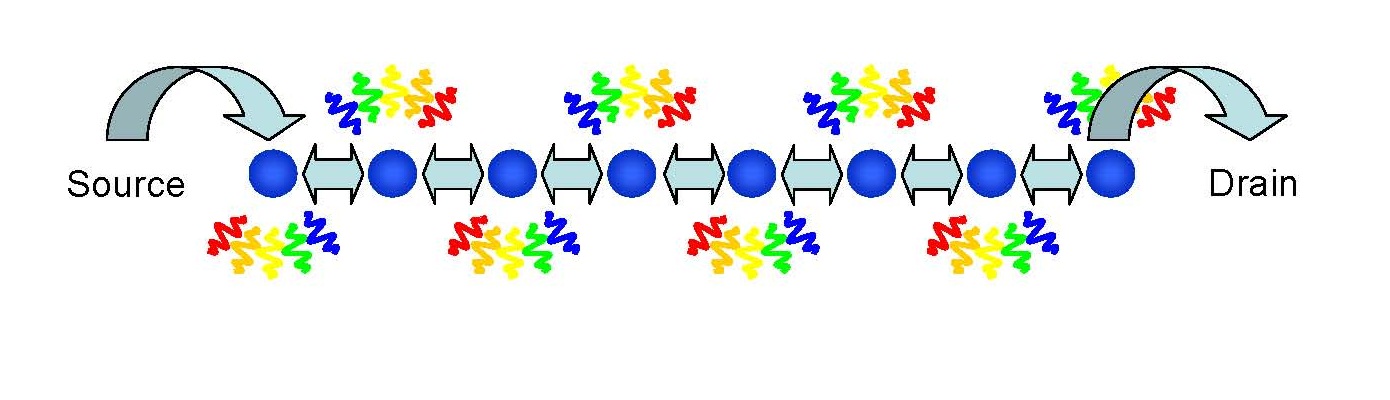}
\caption{A source drives excitations into the first site of
a chain where nearest neighbors are coupled by a hopping
interaction. The excitations leave the chain via the last
site.}
\end{figure}
Besides the coherent hopping between the sites there may also
be thermally activated transitions between nearest neighbours.
These would again be described by Lindblad operators of the
form
\begin{eqnarray}
        {\cal L}_{th}(\rho) \!\!&=& \!\!\sum_{k=1}^{N-1} \Gamma_{k}^{th} \left[
        2\sigma_{k+1}^{+}\sigma_{k}^{-}\rho \sigma_{k}^{+}
        \sigma_{k+1}^{-} - \{ \sigma_{k}^{+}\sigma_{k+1}^{-}
        \sigma_{k+1}^{+}\sigma_{k}^{-},\rho \}\right]\nonumber\\
        \!&+& \!\!\sum_{k=1}^{N-1} \Gamma_{k}^{th} \left[
        2\sigma_{k+1}^{-}\sigma_{k}^{+}\rho \sigma_{k}^{-}
        \sigma_{k+1}^{+} - \{ \sigma_{k}^{-}\sigma_{k+1}^{+}
        \sigma_{k+1}^{-}\sigma_{k}^{+},\rho \}
        \right].\nonumber \\
\end{eqnarray}
Hence the most general form of quantum dynamical equation that
we will be considering here is of the form
\begin{equation}
        \frac{d\rho}{dt} = -\frac{i}{\hbar}[H,\rho] + {\cal L}_{s}(\rho)
        + {\cal L}_{d}(\rho) + {\cal L}_{th}(\rho)
        + {\cal L}_{deph}(\rho).
\end{equation}
To determine the conductivity of such a system we will consider
$p_{sink}(t)$ which may be obtained directly by integrating the
population of the last site multiplied with the net transfer from
that site into the sink, that is $$p_{sink}(t) = \int_{0}^{T}
2\Gamma_{d} \rho_{n,n}(t) dt.$$ As the current we will then define
the asymptotic rate of growth of $p_{sink}(t)$ that is
\begin{equation}
        I = \lim_{t\rightarrow\infty} \frac{dp_{sink}}{dt}(t).
\end{equation}

{\em Quantum coherence \& conductivity --} Now we would like to
demonstrate that
the conductivity of an externally driven channel exhibits resonances
at which the conductivity is strongly suppressed. We will then
show that the depth of these resonance reduces in the presence
of dephasing noise and that it can be related to the amount of
quantum coherence in the channel. Thus the depth of resonances
may be taken as a measure for the presence of quantum coherence.

We consider a channel subject to a constant and a time dependent
potential due to applied electric fields. Furthermore we assume,
for simplicity and clarity of the argument, that the inter-site
hopping rates are constant across the chain, i.e. $c_k=c$. In
this case the coherent part of the dynamics is given by a Hamiltonian
of the form
\begin{equation}
        \label{Ham}
        H/\hbar = \sum_{k=1}^N (\Omega_0 + \Omega_1\cos\omega t)k \sigma_k^{+}\sigma_k^{-} +
        \sum_{k=1}^{N-1} c (\sigma_k^{+}\sigma_{k+1}^{-}
        + \sigma_k^{-}\sigma_{k+1}^{+}).
\end{equation}
Here $\hbar\Omega_0$ is the energy difference between adjacent
sites. In the ion channel this would be related to the energy
difference between two close lying energy levels in adjacent
potential wells. This difference may not exceed the level spacing
in one of the potential wells which in turn tends to be below
$10^{12}s^{-1}$. To achieve a transfer rate of the order of
$10^8s^{-1}$ through the channel, the effective hopping rate
between sites should be of the same order which in turn implies
that $c^2/\Omega_0 \sim 10^8s^{-1}$ if $c\ll \Omega_0$ as is
the case for an ion channel.

In the limit of very long, ideally infinite chains, this Hamiltonian
is known to exhibit the quantum coherent phenomenon of dynamic
localization or dynamical suppression of tunneling,
\cite{DunlapK86,HolthausH96}. The existence of this effect and
its coherent character may be inferred from the following arguments.
Consider a transformed picture in which the time-dependent
on-site energies
vanish at the expense of time dependent coupling strengths between
the neighboring sites \cite{HolthausH96}. This is achieved by
defining $A(t) = -\Omega_0 t - (\Omega_1/\omega) \sin(\omega t)$
to move to an interaction picture via the transformation
$|\tilde{\psi}(t)\rangle =
e^{-i A(t) \sum_k k\sigma_k^{+}\sigma_k^{-}}|\psi(t)\rangle$.
In this interaction picture the dynamics is governed by a
Hamiltonian
\begin{equation}
        H_I/\hbar = \sum_{k=1}^{N-1} c (e^{-i A(t)}
        \sigma_k^{+}\sigma_{k+1}^{-} + e^{i A(t)}
        \sigma_k^{-}\sigma_{k+1}^{+})
\end{equation}
where now the coupling rates are time-dependent.
For small $c$ the hopping dynamics between neighbouring
sites is slow compared to the time dependence $e^{\pm i A(t)}$
and averaging this Hamiltonian over the interval $[-\pi/\omega,\pi/\omega]$
we find that the effective Hamiltonian takes the form
\begin{equation}
        \label{effective}
        H_I/\hbar = \sum_{k=1}^{N-1} c
        J_{\frac{\Omega_0}{\omega}}(\frac{\Omega_1}{\omega})
        (\sigma_k^{+}\sigma_{k+1}^{-} + \sigma_k^{-}\sigma_{k+1}^{+}).
\end{equation}
Hence we expect that if $\Omega_0=n\omega$ for some $n\in \mathbb{R}$
and $E_1/\omega$ coincides with a zero of the Bessel function $J_n$,
then the evolution of a wave-packet becomes periodic in time and the
spreading of the wave-packet and hence transport is suppressed.
For an infinite chain this can be demonstrated rigorously
\cite{DunlapK86,HolthausH96}.

\begin{figure}[htb]
\includegraphics[width=8.5cm]{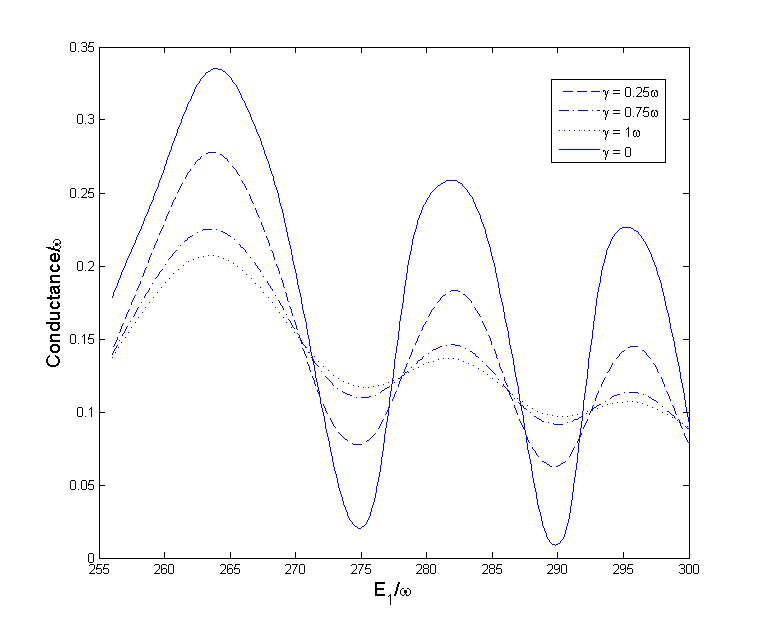}
\caption{\label{figsim} Conductivity versus strength $E_1$ of the
applied ac-potential for a chain with parameters $N = 5$,
$\Omega_0=256\cdot 10^8 s^{-1}$, $\Gamma_{d}= \Gamma_s =  10^8s^{-1}$,
$\omega =  2\cdot 10^8 s^{-1}$, $n_T=1$ and $c=8\cdot 10^8s^{-1}$.
For increasing dephasing rates $\gamma = 0, 5\cdot 10^7s^{-1},
1.5\cdot 10^8s^{-1}$ and $2\cdot 10^8s^{-1}$ one observes
a reduction in the depth of conduction
resonances. }
\end{figure}

For a finite system that is in contact with a source and a sink,
excitation transport will not vanish exactly.
Firstly, within a time interval $[-\pi/\omega,\pi/\omega]$
the wavepacket will not be stationary but will oscillate changing its
width. The localization length (the maximal extent over which the wavepacket
may spread in that interval) may exceed the length of the channel
and hence leave it into the sink. On the other
hand the quantum coherence responsible for the localization will
be perturbed by the incoherent processes that are taking place at
source and sink.  Nevertheless, it can still be expected that
transport is suppressed significantly, even for short chains.
Indeed, a numerical analysis confirms the presence of dynamic
localization even for the dynamics of a short chain. Fig. \ref{figsim}
indeed demonstrates that with just $5$ sites (similar results may
be observed for shorter chains accompanied by higher oscillation
frequencies $\omega$), high contrast resonances in the conductivity
can be observed and that these coincide very well with the zeros
of the relevant Bessel-functions.

Note that the effective coupling rates in eq. (\ref{effective})
emerge through averaging over transition amplitudes and it hence
becomes clear that the suppression of transport is a coherence
effect.
Another way to understand these results is from the observation
that the ac-field will let the onsite energies oscillate so
that periodically in time the neighboring sites become energetically
close enough so that a coupling $c$ that is insufficient to bridge an
energy-gap of the order of $E_0$ will then induce transitions.
In fact, this is the periodic analogue of the mechanism by which
dephasing helps to overcome energy gaps \cite{CarusoCDHP09}.
However, in the periodic case, the wrong timing, can lead to
coherent suppression of transport as well \cite{example}.
The physical origin of this trapping, and its
quantum nature, may be understood by considering the extreme
case of the Hamiltonian
$H/\hbar = \sum_{k=1}^{N-1} \frac{c}{2}[1+sign(sin(\omega t
+ k\pi))](\sigma_k^{+}\sigma_{k+1}^{-} + \sigma_k^{-}
\sigma_{k+1}^{+})$ that is the couplings are switched on and
off with a frequency $\omega$ alternatingly between neighboring
pairs of sites.
%
Within this model it is now easy to see why time-varying coupling
rates may suppress transport in a quantum setting. For $\omega = c$
we observe that
$\exp^{-iH_{eff}\pi/\Omega}|1\rangle = -|1\rangle$
and as a consequence excitations do not propagate at all.
Hence, we expect that for certain frequencies transport will
be suppressed \cite{remarkII}.

All these arguments show that the dynamic suppression of
transport relies on the existence quantum coherence, either
through the destructive interference of transition amplitudes
or the existence of full oscillations between sites. In purely
classical rate equation models or strongly decohered quantum
systems these effects will not exist.
There is no destructive interference nor can the population
between neighboring sites oscillate perfectly as it will
rather approach an equilibrium value in which both sites will
be populated. As a consequence we will not observe complete
suppression of transport in a classical rate equation models
with time-modulated coupling rates between sites. In this
sense dephasing may in fact assist the transport as it suppressed
destructive interference at the resonance points.

The expected sensitivity of this phenomenon to the presence of
decoherence can be seen in Fig. \ref{figsim}. Decreasing depth
of the conductance resonances is associated with increasing levels
of decoherence which in turn can be expected to be associated
with decreasing levels of quantum coherence in the system. This
is indeed a crucial observation as it allows us to turn the
situation on its head and {\em use the depth of the contrast of
the resonances as a tool to determine quantitatively the presence
or absence of quantum coherence in the system} from measurement
of its conductivity.

\begin{figure}[htb]
\includegraphics[width=9cm]{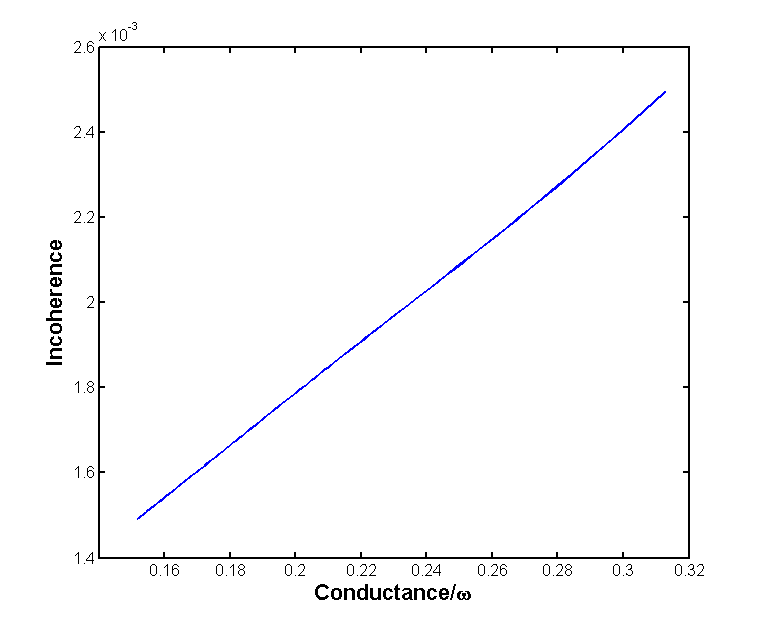}
\caption{\label{cohcon} Incoherence versus conductivity at the
first resonance of fig. \ref{figsim}. The best fit to the
data suggest that incoherence is proportional to conductivity. }
\end{figure}

There is no unique measure for the level of quantum coherence or
incoherence in a state but the observed direct relationship hold
true for a variety of measures. As a measure of incoherence,
the quantity ${\cal C} = \sum_{k\neq l}
|\rho_{k,k}\rho_{l,l}-\rho_{k,l}\rho_{l,k}|$ has the desirable
properties that it vanishes for pure states, increases with
decreasing size of the diagonal elements and is maximized for
maximally mixed state. In addition ${\cal C}$ is concave, that is,
it increases under mixing of quantum states.
For the first resonance in Fig. \ref{figsim}, comparing the
depth of the conductivity resonance with the depth of the
resonance in the quantum incoherence ${\cal C}$ for different
levels of dephasing noise, we find Fig. \ref{cohcon}. This
result is very well fitted by ${\cal C}\sim I$.
%
It is this relationship that we would like to exploit for the
detection of quantum coherence from the observation of currents
in a modulated ion channel.

{\em Theory Discussion \& Speculation --}
The results above demonstrate that systems that are governed by
coherent quantum dynamics will, for certain choices of parameters,
exhibit resonances in their transport
efficiency, i.e. conductivity. As these resonances are absent in
purely classical rate equation models or in quantum systems subject
to very strong decoherence, the observation of currents, for example
in periodically driven ion channels, provides an interesting alternative
for the observation and verification of the presence of quantum coherence.

These resonances, caused by quantum coherence, are relatively sharp
features in parameter space. As a consequence, in the presence of
quantum coherence, the dynamics is much more sensitive to small
variations of system parameters such as ion mass, ion radius etc.
In purely classical models the rather small differences in diameter
between potassium and sodium ions are not expected to result in significant
changes in the ion conduction. In addition to the effects on the
conduction rate presented above one may speculate, whether the
enhanced sensitivity of the quantum dynamics may contribute to the
selectivity of the filter of the ion channel.

Needless to say, the dynamics of ion conduction in ion channels is
in fact a many-body phenomenon in which many degrees of freedom
interact. Recent numerical simulations of the ion-transfer
\cite{Gwan2007} suggest for example that the multi water-potassium
chain does not move as one unit but rather in discrete units of
individual water-potassium translocations, in which the
translocation of a potassium ion to the next site is
followed by the water molecule in picosecond time scale. Given
the short time scale for this substructure of the transport
process it provides an additional mechanism through which
quantum coherence may appear.

Of course such dynamics is considerably more complex than the
model Hamiltonians described here more rigorous conclusions in
that direction need to await more detailed simulations and
experimental investigations \cite{IonChannels}.


%
%

{\em Experimental realization --}
As mentioned earlier, it is an advantage of the model presented
above that it provides experimentally testable predictions for
the existence of coherent excitations in the selectivity filter
through the observation of driven quantum resonances. This is
experimentally particularly attractive as the presence of coherence
can be inferred via the measurement of bulk signals, eliminating
some of the challenges associated with alternative and more direct
methods such as 2D femtosecond spectroscopy. As shown
in Fig.\ref{figsim} in the presence of coherence at certain
frequencies of an external driving field, the ion conduction
through the selectivity filter can be highly impaired. This would
be in sharp contrast to what should be observed when dephasing
is destroying all coherence and the conduction of ions is
governed by pure rate processes making it essentially independent
of the amplitude of the alternating external driving force.

There might be several ways for testing this prediction experimentally.
In the following we would like to discuss one of these
in some detail to demonstrate the in-principle feasibility of such
measurements.
One could for example be using state of the art
electrophysiology methods with some modifications for adapting
them to the particular conditions that are required for resonances
in ion channels to be observed. Electrophysiology techniques such
as patch clamping \cite{Sakmann1995} have found applications in a large number
of biophysical studies for recording single ion channel currents.
During the open gating period of a few $ms$ these currents
can be observed experimentally in different patch clamp techniques
such as the cell attached mode and the excised patch configuration
\cite{Sakmann1995}. In these methods a glass pipette with an inner
diameter of
$~1 \mu m$ is filled with an electrolyte and is used to make a high
resistance ($>10^{9}\Omega$) seal with the cell membrane. Generally, only a
few ion channels can be present on a membrane patch of that size.
Because of the high resistance of the seal any exchange of ions
between the electrolyte in the pipette and the bath (or the cell
cytoplasm) passes through the ion channels. Voltage-gated channels have single-channel conductances typically  range from $10-100 pS$. They last over the few $ms$ open period of the gate are amplified by a current to voltage converter and recorded.

Here we should mention that the fact that multiple ion channels could
be present on a membrane patch of the size of pipette tip by no means
affects the experimental signature of quantum coherence in a negative
fashion. As discussed above the manifestation of quatum coherence in
the selectivity filter would be observable as a change in the "bulk"
ionic current through the channel as a function of the externally
applied AC driving force. In that sense as long as the driving force
is constant over the number of channels, it actually represents an
advantage to have more channels within the patched area, as it would
help in increasing the signal to noise ratio.

But in order to be able to apply electrophysiology methods to
experimentally drive the resonances in the selectivity filter
and record the transmitted ion rate several conditions have
to be met. These will be discussed in some detail in the
remainder of this work.

In contrast to many single ion channel studies in which investigators
are interested in the gating dynamics of the channel, our recordings
have to be made during the open state of the channel. This requires
the reliable preparation of the channel in its open gate state. Since
the structure of the selectivity filter is highly conserved across the
potassium channel family the experiments can also be done in Kv voltage
gated potassium channels \cite{Yellen2002}. For this category of potassium channels a number of pharmacological agents are available which can be used to initiate the open state of the channel.
Alternatively, a class of potassium channels, the calcium
activated potassium channels \cite{Yellen2002, Jiang2002, Schumacher2001}
can be used in which the open conformation can be trigged by adding calcium to the bath. Finally,
it is also possible to introduce site mutations in the voltage
sensor region or use enzymes which impair the function of
the voltage sensor and lead to permanently open channels \cite{Hoshi1990}.

Next, it is necessary to establish an ionic current across the
channel in the absence of the AC driving force. Again, while
this could be done through a current injection via the patch
clamp amplifier, it would be more convenient and also reassemble
the natural conditions by establishing a potassium concentration
gradient across the ion channel. The role of the potassium channel
during the action potential is to restore the membrane potential
from $\sim +30mV$ after the fast opening of the sodium channels back
to the resting membrane potential of $\sim -70mV$. During this period
potassium flows out of the cell along its concentration gradient
($~90mM$ intracellular vs. $~3mM$ extracellular). Given the
equilibrium potential \cite{Hille2001} of potassium of $~-85mV$ at this
concentration difference this represent a driving force of $~115mV$.
For our experimental considerations it would not be necessary to
hold the membrane potential at $+30mV$. Once the potassium
concentration difference ($~90mM$ intracellular vs. $~3mM$
extracellular) is generated by adding more potassium to the
patch pipette than to the bath, even at 0mV membrane potential,
a driving force of 85mV for the ions would be available. If we
assume an excised patch configuration in which a the current through
a few channels on a dissociate patch of membrane sealed to the
pipette is measured, this driving force will establish a constant
current of a few $pA$. This constant driving force and the
associated ionic current represent $ E_{0}$ in equation (\ref{Ham}).

In order to observe the predicted quantum resonances via the
ionic currents through the channel an alternating driving
potential, $E_{1}\cos\omega t$ with the frequency $\omega$
has to be applied across the channel. This frequency $\omega$
should be chosen to exceed the effective hopping rate in our
theoretical model which in turn is expected to be of the order
of the ion channel transmission rate, i.e. around $10^{8}s^{-1}$.
As argued before, for an energy difference $\hbar\Omega_0$
between adjacent potential wells, the resonances will be
found at the zeros of the Bessel function $J_{\Omega_0/\omega}$.
We can expect $\Omega_0\gg \omega$ and hence the first zero
of the Bessel function $J_{\Omega_0/\omega}$ is, in leading order,
located at $\Omega_1/\omega \cong \Omega_0/\omega$
\cite{Jeffreys}.
The required ac-field could be applied
across the channel by using an external function generator
synchronized with the amplifier. Alternatively an RF field
at the same frequency with its polarization aligned along
the axis of the ionic propagation might be used as an
alternating driving force.

\begin{figure}[t]
\includegraphics[width=8 cm]{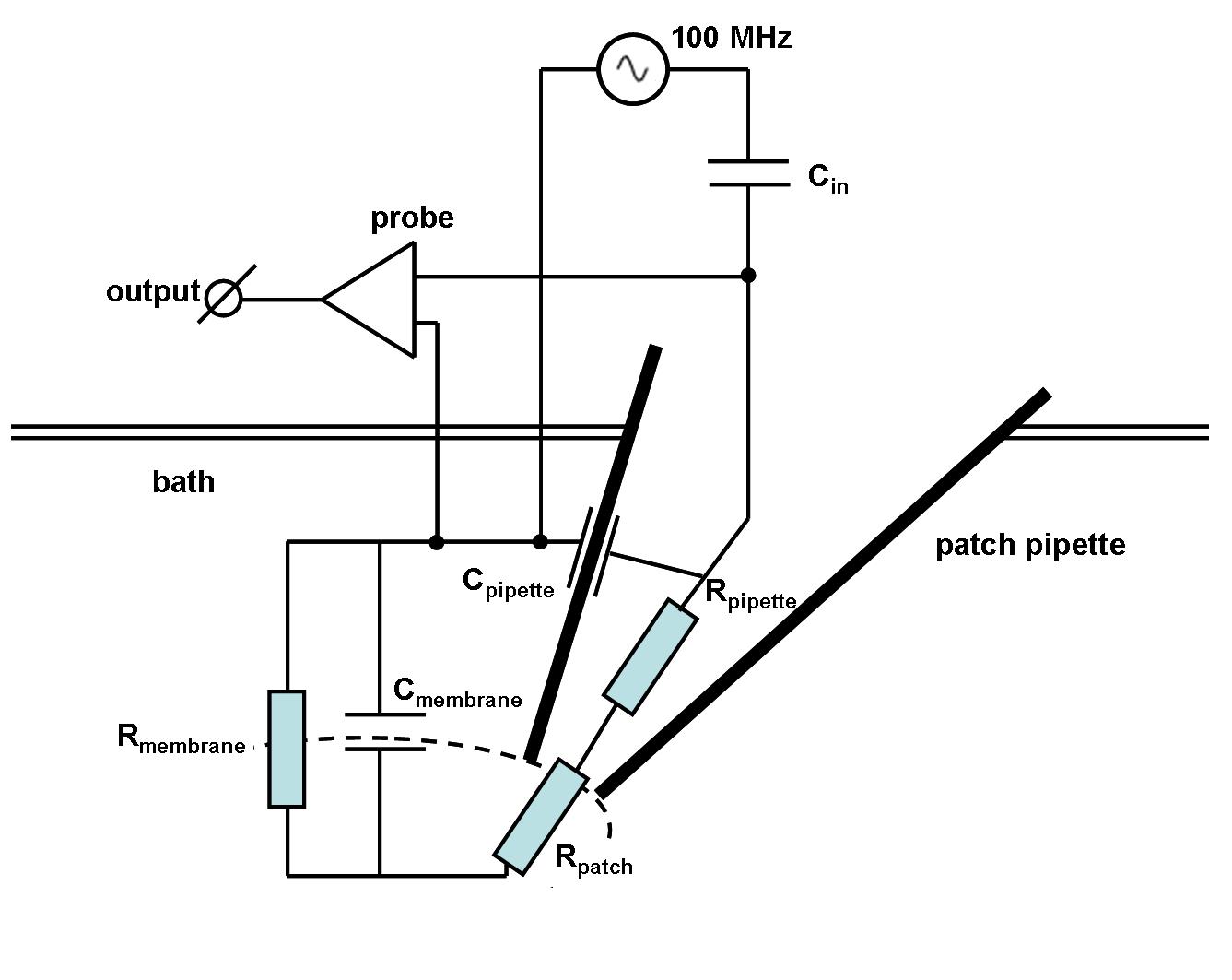}
\caption{\label{patch}Equivalent cuircuit for the excised patch
configuration and external driving force to induce quantum resonances.}
\end{figure}

However, because of the high frequency of the AC field the
membrane capacitance and the stray capacitance of the pipette
glass will lead to additional capacitive currents significantly
higher that the membrane current. Moreover, the time constant
$\tau$ for establishing a potential difference across the membrane
is given by $\tau=C_{m}R_{p}$, with $C_{m}$ the membrane capacitance
and $R_{p}$ the pipette resistance. Therefore it would be desirable
to minimize the area of the membrane patch across which the AC
field is applied. This limitation excludes any whole cell recordings
and limits our considerations to the excited patch configuration
for which the membrane capacitance is minimized as it will be
determined by the surface area of the pipette tip. (see
Fig \ref{patch}). The excised patch configuration is also
advantageous as it allows the study of the ion channels in
isolation and provides better control for the application agents
on both sides of the channel.




Assuming an inner diameter of $1\mu m$ and a typical membrane
capacitance of $C_{m} \sim 1 \mu F/cm^{2}$ this results in a membrane
capacitance of $C_{m}\sim 8 \times 10^{-15}F$. As we would like to
apply an AC field at $~100MHz$ the time constant $\tau$ should be
at least $~ 5ns$. Based on the membrane capacitance of $C_{m}\sim 8
\times 10^{-15}F$ estimated above, it means that the pipette
resistance $R_{p}$ should not be larger than $0.6 M \Omega$.
Typical pipette resistances are in the range of tens of $M \Omega$,
but it should be possible to reduce these values by appropriate
shaping of the pipette tip.

Further, the mentioned capacitive currents required some careful
considerations. As shown in Fig \ref{patch} the membrane capacitance
$C_{m}$ and the pipette capacitance $C_{p}$ are both contributing
to the induced capacitive currents by the fast alternating AC field.
While the membrane capacitance can be minimized by using the excised
patch configuration, minimizing the pipette capacitance requires
using special type of glasses \cite{Odgen1994}, minimizing the level of the
electrolyte in the pipette and coating the pipette with insulating
raisins which would thicken the external pipette walls and reduce
the stray pipette capacitance. This can be done by coating the
pipette with a layer of Sylgard, an inert, hydrophobic, translucent
elastomer raisin which is cured rapidly by heat \cite{Rae1992}.
These measures could reduce the pipette capacitance to
$C_{p}\sim 10^{-13} F$. However, this would be still about an order
of magnitude larger than $C_{m}$ and therefore the major contributor
to the capacitive current.

As discussed above the applied AC field has to induce a membrane
potential change of $~ 315mV$ in $~10ns$. With $C_{p}$ as the
dominating factor this means that the AC field would lead to a
capacitive current of $~ 3 \mu A$ which is six orders of magnitude
higher that the $pA$ ionic currents through the channel. It could
be somewhat challenging to separate these two components, however
the difference in the timing of these currents could provide the
means to separate them. The above choice for the time constant
$\tau$ is such that the peak the peaks of the capacitive currents
are separated by the period of the AC field. Therefore a lock in
detection technique can be used to reject the capacitive currents
and isolate the ionic current. Moreover for a fixed frequency
(e.g. $100MHz$) it is possible to design additional electronic
circuits which introduce negative capacitance to compensate the
capacitive currents. Finally a number spectral analysis technique
can be applied to isolate the ionic current from the capacitive
currents. Another approach which could eliminate the pipette
capacitance altogether is to use chambers with suspended planar
phospholipids bilayers in which the ion channels have been
reconstituted \cite{Miller1986}. In this technique the two macroscopic
chambers filled with electrolytes are separated by an artificial
lipid layer in which the ion channels have been reconstituted.
Ions can move along their concentration gradient only through
the ion channels and are recoded and amplified with a same
amplifier system as discussed above. The AC field can be applied
between the two chambers, while the current through the membrane
is monitored. Since in this approach the pipette is eliminated
the only source for capacitive currents is the lipid membrane.
However in such a setup the membrane area is typically larger
than the in the patch pipette so that it might lead only to a
moderate reduction of the capacitive currents.

Finally, the recent development of NV-centers point to an alternative
method for recording pico-Ampere currents in biological systems
with high spatial and temporal control \cite{Wrachtrup}.

On the positive side the fact that the flux rate of ions at a
given concentration is fairly constant and the fact that the
width of the expected quantum resonances in Fig. \ref{figsim}
is not very sharp, is expected to help finding these resonance
experimentally.

Overall the above assessments show that currently available
techniques from single channel biophysics studies with moderate
modification can be adopted and used for experimental investigations
of quantum resonances in the selectivity filter of potassium channels.

{\em Acknowledgements --} This work was supported by an Alexander von Humboldt-Professorship and by the Howard Hughes Medical Institute.
Correspondence should be adressed to vaziria@janelia.hhmi.org or to
martin.plenio@uni-ulm.de.


\begin{thebibliography}{99}
\bibitem{Estimate} To estimate the expected decoherence time we
assume that the dominant form of noise stems from the stretch
mode of the carbonyl groups in the ion channel. Their resonance
frequency is approximately $\omega_{CO} \sim 3.2\cdot 10^{14}s^{-1}$
corresponding to a wave number of around $1700cm^{-1}$. Hence, at
room temperature, this oscillation is essentially in the ground state.
This zero point motion gives rise to a mean square deviation of the
position of the oxygen atoms of the order of $0.02$\AA. The oxygen
atoms in the carbonyl group carry an effective negative charge and
hence their motion will affect the potential that charges such
as potassium ions experience in the ion channel. The fluctuating
potential leads to fluctuating energy differences between two
potential wells which in turn are responsible for dephasing of
superposition states between two wells. The potassium ions in
their potential wells have a frequency of around
$\omega_{K}\sim 3\cdot 10^{12}s^{-1}$. Numerically one can now determine
that the zero-point motion of the carbonyl groups will affect the
energy difference between two wells by around $\hbar \omega_{K}/10$
for a typical displacement. This leads to a coupling of the oscillations
of the carbonyl groups to that of the potassium ions. Now we treat this
as a simple two-mode quantum system, one mode referring to the
oscillation of the carbonyl group and the other to the energy difference
between two potential wells. The Hamiltonian is given by
$H = \hbar\omega_{CO}a^{\dagger}a + \hbar\omega_{K}(X_{CO})
b^{\dagger}b \cong \hbar\omega_{CO}a^{\dagger}a +
0.1\hbar\omega_K(a+a^{\dagger})b^{\dagger}b+
\hbar\omega_K b^{\dagger}b$. Here $\hbar\omega_K$ denotes the
energy difference between two potential wells which is in turn
a function of the position of the oxygen atom in the carbonyl
group. Now initialising the system in the state
$|0\rangle_a(|0\rangle+|1\rangle)_b$ corresponding to a superposition of
an ion in the left or right potential well we find that the
time in which the fidelity with the initial state has dropped
to $1/e$ is around $3000 \omega_K^{-1}$ suggesting that the
decoherence time is of the order of
$t_{dec}\sim 3000 \omega_K^{-1} \sim 3\cdot 10^{-9}s$. This
is of the order of the tunneling time. It should be noted that
there are a variety of unknowns that may affect the decoherence
time. On the one hand we have ignored the oscillation of the
carbon, nitrogen and hydrogen in the ion channel which may lead
to a slightly higher decoherence rate. We have
neglected oscillations along the channel axis as these have a
much smaller effect on the potential and hence couple more
weakly to the potassium ions.  On the other hand the present
estimate assumes a first order effect of the oscillation of the
carbonyl group on the potential in the ion channel. If the
system was making oscillations around equilibrium the potential
change would be a second order effect and hence smaller. Hence
the quoted figure provides a ball park figure but may vary
somewhat. Furthermore, it should be noted
that these oscillations do not only lead to a dephasing
between two potential wells but also affects the height of
the potential barrier. In fact, the line width $\Delta \omega$
due to the oscillations and the variations in the height
of the potential barrier are approximately proportional.
As the tunneling rate is strongly non-linear with the potential
barrier height it can be expected that these oscillations
can actually be beneficial.
\bibitem{example} Consider
$H/\hbar = \sum_{k=1}^{N-1} \frac{c}{2} [1 + a\cdot
sin(\omega t + k\pi)] (\sigma_k^{+}\sigma_{k+1}^{-} +
\sigma_k^{-}\sigma_{k+1}^{+})$
where $0\le a\le 1$ and the coupling rates between neighbouring
pairs oscillate with opposite phase. The resulting Hamiltonian
can be solved exactly in the limit of $N\rightarrow\infty$ as
it is invariant by a shift of two sites and exhibits dynamic
localization. Indeed, an excitation initially localized in site
$1$ will not propagate through the system completely but a portion,
depending on the size $a$, is trapped depending.
\bibitem{remarkII} Note that for the case $\omega = 2c$ we observe
a certain enhancement of the transport because
$\exp^{-iH_{eff}\pi/\omega}|1\rangle = i|2\rangle$
so that at the end of the interval $[0,\pi/\omega]$ the population
has been transferred perfectly to site $2$. In the next time
interval sites $2$ and $3$ are now coupled with the same strength
and as a consequence the population is transferred perfectly to
site $3$ and so forth. Hence the population is transported without
any spreading and at the maximum speed given the coupling rate
$c$ through the chain.
\end{thebibliography}
\end{document}